\documentclass[journal]{IEEEtran}
\usepackage[colorlinks,linkcolor=blue,anchorcolor=blue,citecolor=blue,bookmarks=true]{hyperref}  
\usepackage[T1]{fontenc}
\ifCLASSINFOpdf
\usepackage{booktabs}
\usepackage{multirow}
\usepackage[pdftex]{graphicx} 
\DeclareGraphicsExtensions{.eps,.pdf,.jpeg,.png}
\else
\usepackage[dvips]{graphicx}
\fi

\usepackage[compress,nospace]{cite}
\usepackage[table,xcdraw]{xcolor}
\usepackage[cmex10]{amsmath}
\usepackage{epstopdf,amsthm,stfloats,siunitx,amssymb,wasysym,algorithm,algorithmic,array,url, color,subfigure} 
\usepackage{footnote}
\makesavenoteenv{tabular}
\usepackage{soul}
\usepackage{caption}
\soulregister\cite7 
\begin{document}

	\title{Low-Cost Beamforming and DOA Estimation Based on One-Bit Reconfigurable Intelligent Surface}
	
	\author{Zihan~Yang, Peng~Chen,~\IEEEmembership{Senior Member,~IEEE},
Ziyu~Guo,~\IEEEmembership{Member,~IEEE}, and Dahai Ni
\thanks{This work was supported in part by the Natural Science Foundation for Excellent Young Scholars of Jiangsu Province (Grant No. BK20220128), the National Key R\&D Program of China (Grant No. 2019YFE0120700), and the National Natural Science Foundation of China (Grant No. 61801112).\textit{(Corresponding author: Peng Chen)}}
\thanks{Z.~Yang and P.~Chen are with the State Key Laboratory of Millimeter Waves, Southeast University, Nanjing 210096, China (email: \{yangzihan, chenpengseu\}@seu.edu.cn).}
\thanks{Z.~Guo is with the School of Information Science and Technology, Fudan University, Shanghai 200438, China (email: zguo@fudan.edu.cn).}
\thanks{D.~Ni is with the Yangzhou Institute of Marine Electronic Instruments, Yangzhou 225100, China (email: seuseayou@outlook.com).}}

	\maketitle
	
	\begin{abstract}
	In this work, we consider the Direction-of-Arrival (DOA) estimation problem in a low-cost architecture where only one antenna as the receiver is aided by a reconfigurable intelligent surface (RIS). We introduce the one-bit RIS as a signal reflector to enhance signal transmission in non-line-of-sight (NLOS) situations and substantially simplify the physical hardware for DOA estimation. We optimize the beamforming scheme called measurement matrix to focus the echo power on the receiver with the coarse localization information of the targets as the prior. A beamforming scheme based on the modified genetic algorithm is proposed to optimize the measurement matrix, guaranteeing restricted isometry property (RIP) and meeting single beamforming requirements. The DOA results are finely estimated by solving an atomic-norm based sparse reconstruction problem. Simulation results show that the proposed method outperforms the existing methods in the DOA estimation performance.
	\end{abstract}
	
	\begin{IEEEkeywords}
		DOA estimation, reconfigurable intelligent surface, genetic algorithm, atomic norm.
	\end{IEEEkeywords}
	
	\section{Introduction} \label{sec1}
	Recently, reconfigurable intelligent surface (RIS) has been widely applied in wireless communication and target localization \cite{yan2020passive,basar2020reconfigurable,wu2019towards} due to its high cost-effectiveness and channel reconstruction ability. Traditionally, high-resolution algorithms are studied to enhance the parameter estimation performance in the DOA system due to the low SNR in the typical non-line-of-sight (NLOS) environment \cite{8663559}. RIS can intelligently reconstruct the propagation environment of electromagnetic waves to a desirable form \cite{dai2020reconfigurable} with several positive intrinsic-negative (PIN) diodes integrated into RIS units. Based on this characteristic, RIS is a more flexible tool for source localization in complex NLOS situations\cite{chen2021reconfigurable,esmaeilbeig2021irs}. Ref. \cite{wan2020deep} utilizes RIS to locate the close target vehicles with relatively high accuracy when obstacles block the GPS signal. Ref.\cite{albanese2021papir} develops a practical RIS-aided localization system called PAPIR to finely estimate the user equipment position when there is no direct link between the user equipment and the access point.
	
	DOA estimation plays an important role in wireless communication, UAVs localization and the Internet of Things \cite{wang2019assistant,huang2018deep,zheng2021augmented}. The unknown channel parameters can also be converted to the solutions of a sparse multidimensional DOA estimation problem\cite{ardah2021trice}. In a RIS-aided DOA estimation model, an additional measurement matrix with RIS phase offsets as elements is the main contribution of the RIS \cite{wagner2021gridless,wei2020gridless}. In \cite{he2021channel}, the measurement matrix has been optimized to maximize the effective SNR at the receiver. In \cite{lin2021single}, the authors construct the measurement matrix with completely random radiation patterns to measure the targets multiple times. The measurement matrix can be seen prior to further estimating DOA. Compressed sensing technologies such as the $\ell_1$-singular value decomposition ($\ell_1-\text{SVD}$) and sparse Bayesian learning algorithms can achieve advanced resolution. And the off-grid error can be reduced when dynamically meshing the spatial or assuming the error obeys some known distribution \cite{wang2019assistant}. The atomic norm minimization (ANM) method avoids grid mismatch by formulating the DOA estimation in the continuous spatial domain\cite{wagner2021gridless,wei2020gridless, he2021leveraging }.

	While the measurement matrix and the corresponding random beam patterns are successfully demonstrated, it becomes challenging to directly obtain the low correlated measurement matrix for beams with the same characteristics, such as fixed pointing and suppressed sidelobes. To our knowledge, there is still no universal optimization scheme for such a measurement matrix.
	
	This paper presents a one-bit RIS-assisted direction-finding system with only one omnidirectional antenna as the receiver. The system is low-cost for only one RF chain. We construct a measurement matrix introduced by the RIS to measure the targets multiple times. All measurement vectors obtained by the proposed optimization scheme will only create a specific beam pattern. The correlations of the matrix will also be limited. Since the RIS only reflects the coarse directional information to the receiver, the fine DOAs will be located by sorting to the solution of an atomic norm-based dual polynomial. Unlike the existing sparse representation schemes that adopt a constant as the regularization parameter, we experiment to explore the relationship between the parameter value and SNR.

	
	\emph{Notations:} Lowercase bold letters represent vectors and uppercase bold letters represent matrices. $\left ( \cdot  \right ) ^{\mathrm{H}}$ is  the Hermitian transpose of a matrix. $\left \| \cdot  \right \| _{2} $ represents $\ell_2$-norm of a vector. $\sigma_{\boldsymbol{a}}$ is the standard deviation of the vector $\boldsymbol{a}$. The covariance of vectors $\boldsymbol{a}$ and $\boldsymbol{b}$ is defined as $\mathrm{cov}\left ( \boldsymbol{a},\boldsymbol{b} \right ) $.

	\section{System Model}\label{sec2}
	\begin{figure}[htbp]
		\subfigure[]
		{
			\begin{minipage}{3.5cm}
				\centering
				\includegraphics[scale=0.5]{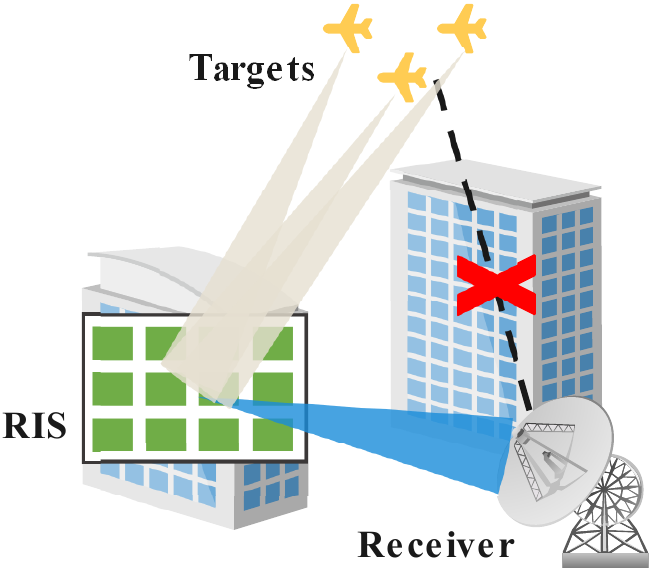}
				\label{aaa}
			\end{minipage}
		}
		\subfigure[]
		{
			\begin{minipage}{5cm}
				\centering
				\includegraphics[scale=0.47]{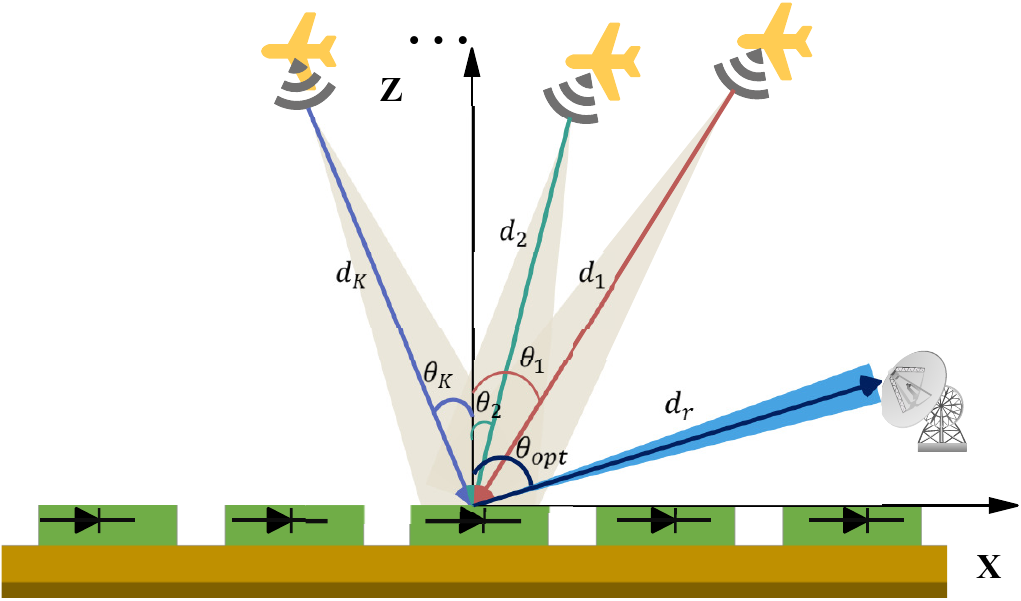}
				\label{bbb}
			\end{minipage}
		}
		\captionsetup{font={footnotesize}}
		\caption{\,\, System architecture model for RIS aided UAVs localization system and the details for DOA estimation. (a) System model (b) Detailed information.}
		\label{system}
	\end{figure}
	
This paper proposes a low-cost direction-finding system with RIS, with only one omnidirectional antenna as the receiver. As shown in Fig.~\ref{aaa}, the system sends a detection signal to the RIS, which reflects the signal to the spatial space. Then, the RIS receives the echo signals from UAVs. The echoes can be reflected in specific directions with the RIS in a designed coding scheme. The direct links between UAVs and the receiver are assumed to be blocked. Multiple measurements of the echoes at the RIS can be realized by controlling the RIS working in a series of coding sequences called the measurement matrix.

The RIS is assumed to be an $N$-element uniform linear array (ULA) with the interval being $d$.  Thus, only the azimuth angles of UAVs are considered in this study; an extension to a planar array can be done. For $K$ far-field UAVs with the direction being  $\boldsymbol{\theta}=\begin{bmatrix}
		\theta_1,\theta_2,\dots,\theta_{K}
	\end{bmatrix}^{\text{T}}$, the signal received by the $n$-th element can be expressed as
	\begin{align}
	x_{n}=\sum_{k=1}^{K}s_{k}e^{j\frac{2\pi}{\lambda}(n-1)d\sin{\theta _{k}}}, n=1,2,\dots, N,
	\end{align}	
where $\lambda$ denotes the wavelength, $\theta_{k}$ is the DOA of the $k$-th signal and $s_k$ is the complex amplitude of the $k$-th echo signal.

In practice, the one-bit controllable configuration with only two phase shifts ($0$ and $\pi$) becomes the most common form of the RIS. The reflection coefficient of each RIS element is assumed to be one. The RIS can reflect incident echo signals several times with multiple coding schemes, which can be seen as multiple measurements of incident signals. The received signal at the receiver corresponding to the $p$-th measurement is given by
	\begin{align}
		y_p= \sum_{n=1}^{N}e^{j\phi_{p,n}}x_{n}+w_p, p=1,2,\dots,P,
	\end{align}
	where $\phi_{p,n}$ is the phase shift caused by the $n$-th RIS element during the $p$-th measurement, and $w_p\in\mathbb{C}$ denotes the zero-mean additive white Gaussian noise with variance being $\sigma^2_{\text{w}}$.

Then, collecting all $P$ measurements, the received signal can be expressed as
	\begin{align}
		\boldsymbol{y} &=[y_1,y_2,\dots,y_P]^{\text{T}}
		\notag \\
		&=\sum_{n=1}^{N}\sum_{k=1}^{K} s_{k}e^{j\frac{2\pi}{\lambda}(n-1)d\sin\theta_k }\begin{bmatrix}
			e^{j\phi_{1,n}} \\
			e^{j\phi_{2,n}} \\
			\vdots \\
			e^{j\phi_{P,n}}
		\end{bmatrix}  
		+\boldsymbol{w}
		\notag \\ 
		&=\boldsymbol{U}\boldsymbol{A}(\boldsymbol{\theta})\boldsymbol{s} +\boldsymbol{w},
		\label{model}  
	\end{align}
	where $\boldsymbol{w} = \left[w_1, w_2,\dots,w_P\right]^{\text{T}}$, and the signal vector is defined as $\boldsymbol{s}= \begin{bmatrix}
		s_{1},s_{2},\dots ,s_{K}
	\end{bmatrix}^{\text{T}}$. 
We define the measurement matrix as
	\begin{align}
		\boldsymbol{U}= \begin{bmatrix} 
			\boldsymbol{u}_1, \boldsymbol{u}_2,\dots,\boldsymbol{u}_P
		\end{bmatrix}^{\text{T}},
	\end{align}
	where $\boldsymbol{{u}_p}$ represents the $p$-th measurement vector
	\begin{align}
		\boldsymbol{u}_p= \begin{bmatrix}
			e^{j\phi_{p,1}},
			e^{j\phi_{p,2}},
			\dots,
			e^{j\phi_{p,N}}
		\end{bmatrix}^{\text{T}}.
	\end{align}
	As mentioned above, the phase shift $\phi_{p,n}$  is either $0$ or $\pi$. 
We also define a steering matrix $\boldsymbol{A}(\boldsymbol{\theta})$ with the directions being $\boldsymbol{\theta}$ as 
\begin{align}
	\boldsymbol{A}(\boldsymbol{\theta})= \begin{bmatrix}
		\boldsymbol{a}(\theta_1), \boldsymbol{a}(\theta_2), \dots, \boldsymbol{a}(\theta_{K})
	\end{bmatrix},
\end{align}
where the steering vector is defined as 
\begin{align}
	\boldsymbol{a}(\theta_k)=\begin{bmatrix}
		1, e^{j2\pi\frac{d}{\lambda}\sin\theta_k},\dots, e^{j2\pi\frac{(N-1)d}{\lambda}\sin\theta_k}
	\end{bmatrix}^{\text{T}}.
\end{align}
The parameters mentioned in (\ref{model}) are marked in Fig.~\ref{bbb}. Our goal is to finely estimate DOAs $\boldsymbol{\theta}$ of the UAVs based on the received signal $\boldsymbol{y}$ on the receiver in which we assume the knowledge of the signal vector $\boldsymbol{s}$ and coarse UAVs locations in front of the RIS. In this model, the multiple coding schemes of RIS, which build up a measurement matrix $\boldsymbol{U}$, is the main factor influencing estimation performance. The proposed optimization scheme for construction is further introduced in the next section.

	\section{Methodologies}\label{sec3}
	We try to design a measurement matrix with multiple measurement vectors. All the vectors can make the RIS reflecting the echo signals to a specific beam direction different from random RIS beams considered in other literature. Simultaneously, these vectors provide a good restricted isometry property (RIP). In this study, the RIS is set towards the targets. Under the normal incidence of plane waves, the far-field function scattered by the RIS is
	\begin{align}
		f(\theta)=\left|\sum_{n=1}^{N}e^{j\phi_{p,n}+j\frac{2\pi}{\lambda}(n-1)d\sin\theta}\right|, 
	\end{align}
	where $\theta$ is the grid angle in the azimuth direction, and the phase shift $\phi_{p,n}$ is $0$ or $\pi$. The normalized form in dB can  be obtained as
	\begin{align}
		F_{\text{dB}}(\theta)=20\log_{10} \frac{f(\theta)}{\max_\theta f(\theta)} .
		\label{pattern}
	\end{align}

	\subsection{Non-convex Measurement Matrix Optimization Approach}
	The optimization formulation for a single beam can be expressed as 
	\begin{align}
		F= \alpha \sum_\theta |F_{\text{dB}}(\theta)-F_{\text{target}}(\theta)|^{2},
		\label{GA_equation}
	\end{align} 
	where $\alpha$ is defined as
	\begin{align}
		\alpha=\max\left\{\frac{l_{\text{th}}}{l_{\text{cal}}},\frac{\left | \theta_{\text{cal}}-\theta_{\text{opt}} \right | }{\theta_{\text{th}}} \right\},
		\label{alpha}
	\end{align}
	where $l_{\text{cal}}$ and $\theta_{\text{cal}}$ denote the highest sidelobe value in dB and the main beam pointing of current calculated reflection pattern, respectively. $\theta_{\text{opt}}$ is the angle of departure of RIS-receiver as shown in Fig.~\ref{bbb}. The two parameters $l_{\text{th}}$ and $\theta_{\text{th}}$ can be seen as the limits at the beam pointing and side lobe levels, respectively.
	
\begin{figure}[htbp]
		
		\begin{minipage}[t]{0.44\linewidth}
		 	\centering
			\includegraphics[width=4cm]{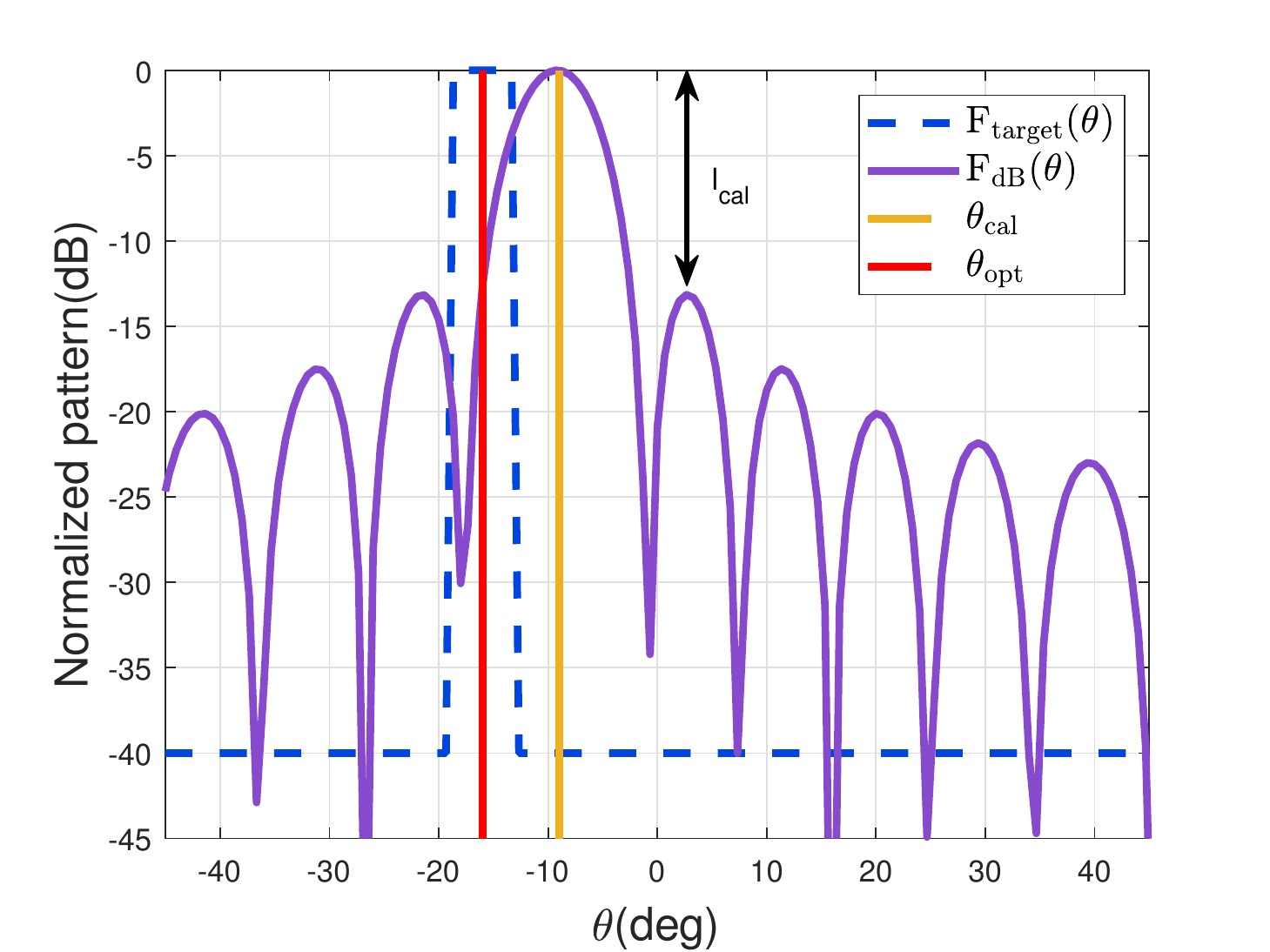}
			\captionsetup{font={footnotesize}}
			\caption{\,\, An example of $F_{\text{dB}}(\theta)$ and $F_{\text{target}}(\theta)$.}
			\label{shili}
		\end{minipage}
		\begin{minipage}[t]{0.5\linewidth}
			\centering
			\includegraphics[width=5cm]{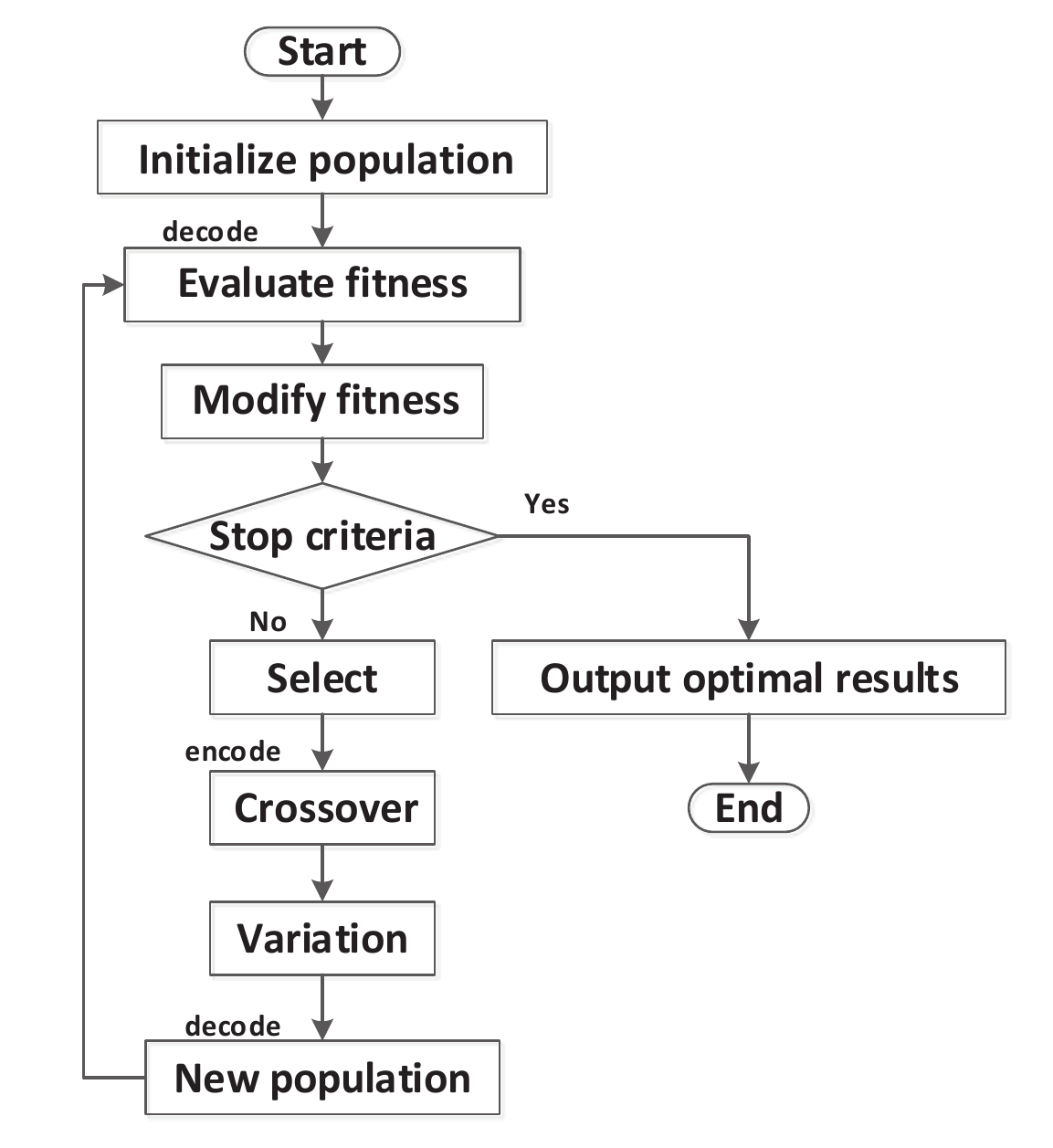}
			\captionsetup{font={footnotesize}}
			\caption{\,\, Flowchart of the modified genetic algorithm.}
			\label{GA}
		\end{minipage}

	\end{figure}

	The construction of $F_{\text{target}}$ is designed to constrain the shape of the reflection beam pattern. As shown in the Fig.~\ref{shili}, the offset between $\theta_{\text{opt}}$ and the $\theta_{\text{cal}}$ enlarges the value of function (\ref{GA_equation}). The beamforming optimization problem with two constraints on both beam direction and the highest sidelobe value given in (\ref{GA_equation}) is non-convex, which is then solved by a modified GA algorithm. The modification is mainly to constrain the mutual coefficient among the measurement vectors.

	To begin with, we establish a mapping between the solution space of the original problem and the bit string space by transforming the phase shift $e^{j0}$ and $e^{j\pi}$ to $0$ and $1$, respectively. A random binary population is initialized with individual size $N_p$ and chromosome size $N$, and the $i$-th individual in the population can be expressed as
	\begin{align}
		\boldsymbol{h}_{i,g},i & = 1,2,\dots,N_p,
	\end{align}
	where the index $g$ means the number of genetic generations. The initial value of each individual accords with uniform distribution. The fitness function of GA algorithm is defined as the reciprocal of (\ref{GA_equation}).
	
	The natural process consists of selection, recombination, and mutation of offspring, which are performed to update the population until an optimal measurement vector is found \cite{mirjalili2019genetic}. Here we adopt the roulette wheel operator to select. The probability of inheritance for each individual depends on the proportion of its fitness to that of all individuals. After completing the given number of cycles, the algorithm terminates.

	The optimum of the GA algorithm can be converted to a measurement vector. This solving procedure is repeated multiple times, generating a measurement matrix to satisfy the constraints. To guarantee successful DOA estimation of $K$ targets in sparse representation problem, the mutual coherence of the matrix or the largest absolute correlation between any two vectors should be less than $\frac{1}{2K-1}$\cite{candes2008introduction}. In the modification term, we calculate the  mutual coefficients among individuals in the population and the obtained measurement vectors. The correlation coefficient between two vectors $\boldsymbol{u}_i$ and $\boldsymbol{u}_j$ is defined as
	\begin{align}
		\rho \left ( \boldsymbol{u}_i,\boldsymbol{u}_j \right ) =\frac{\mathrm{cov}\left ( \boldsymbol{u}_i,\boldsymbol{u}_j \right )  }{\sigma _{\boldsymbol{u}_i}\sigma _{\boldsymbol{u}_j} }.
	\end{align}
	Then we reduce the fitness value of a highly correlated individual by multiplying it with a small value while the rest remains unchanged. The whole flowchart of the modified GA algorithm is shown in Fig.~\ref{GA}.

	\subsection{Atomic Norm-Based DOA Estimation Method}
	The process of DOA estimation can be concluded in three steps. Firstly, suppose we know the prior information that the uncontaminated received signal $\boldsymbol{\hat{y}}$ from $\boldsymbol{y} = \boldsymbol{\hat{y}+w}$ can be written as a non-negative linear combination of a few atoms from the atomic set $\mathcal{A}$, where $\boldsymbol{\hat{y}}=\boldsymbol{UAs}$. The atomic set is defined as
	\begin{align}
		\mathcal{A}=\left \{ \left [ 1,\,e^{j\pi\sin\theta}\,\cdots\,,e^{j\pi(N-1)\sin\theta} \right ]^{\text{T}} , \theta\in (-\frac{\pi}{2},\frac{\pi}{2}]\right \} .
	\end{align}
	The set $\mathcal{A}$ can be viewed as an infinite dictionary indexed by the continuously varying parameter $\theta$. Then, the atomic norm of $\boldsymbol{b}$ can be expressed as
	\begin{align}
		\|\boldsymbol{b}\|_{\mathcal{A} }=&\inf \bigg \{
		\sum_k c_k: \boldsymbol{b}=\sum_k c_k e^{j\vartheta_k}\boldsymbol{a}(\theta_k),\notag\\&
		c_k>0, \vartheta_k\in[0,2\pi),\theta_k \in (-\frac{\pi}{2},\frac{\pi}{2}]
		\bigg\},
	\end{align}
	where $c_k$ is the non-negative coefficient of $k$-th selected atom and $\vartheta$ is the corresponding phase. 
	Secondly, as for the atomic norm denoising method, we can characterize the performance of the estimate $\boldsymbol{\hat{y}}$ from $\boldsymbol{y}$ that solves
	\begin{align}
		\min_{\boldsymbol{x}}\frac{1}{2}\left \| \boldsymbol{Ux}-\boldsymbol{y} \right \|_{2}^{2}+\epsilon\left \| \boldsymbol{x} \right  \|  _{\mathcal{A}}, 
		\label{op}
	\end{align}
	where $\epsilon > 0$ is a parameter related to noise level\cite{chi2020harnessing}. The dual problem of (\ref{op}) is
	\begin{align}
		\max_{\boldsymbol{x}}\quad& -\frac{1}{2}\left \| \boldsymbol{y}-\boldsymbol{Ux} \right \|_{2}^{2}  
		\notag\\s.t.\quad &\left \| \boldsymbol{x} \right \| _{\mathcal{A} }^{\ast } \le \epsilon,
	\end{align}
	which is equal to
	\begin{align}
		\min_{\boldsymbol{x}\in \mathbb{C}^{N}} \quad &\left \| \boldsymbol{y}-\boldsymbol{Ux} \right \|_{2}^{2}  
		\notag\\s.t.\quad&\left \| \boldsymbol{x} \right \| _{\mathcal{A} }^{\ast } \le \epsilon,
		\label{cvx}
	\end{align}
	where $\left\| \boldsymbol{x} \right \| _{\mathcal{A} }^{\ast }=\max_{\boldsymbol{b}\in \mathcal{A}} \left | \boldsymbol{x}^{\mathrm{H}}\boldsymbol{b} \right |$ is the dual atomic norm. To solve the problem in (\ref{cvx}), we use CVX, a package for specifying and solving convex programs utilizing the interior-point methods \cite{grant2014cvx}. Hence, the computational complexity is approximately  $\mathcal{O}\left ( N^{3.5} \right ) $ \cite{wang2018ivdst}. We denote the optimal solution in (\ref{cvx}) as $\hat{\boldsymbol{x}}$. Therefore,  DOAs can be resorting to the peaks of spectrum constructed by the dual polynomial $z(\theta)=\left \langle \boldsymbol{Ua}(\theta),\hat{\boldsymbol{x}}   \right \rangle $.
	\section{Simulation Results}\label{sec4}
	
	This section gives simulations to verify the feasibility of the proposed optimization scheme and DOA estimation method. The Matlab code for the proposed method is available online (\url{https://github.com/chenpengseu/Beamforming-DOA-RIS.git}). The simulation parameters are listed in Table \ref{tab1}. The two parameters $l_{\text{th}}$ and $\theta_{\text{th}}$ in (\ref{alpha}) are -5 and 5, respectively.
	\begin{figure*}[htbp] 
		\begin{minipage}[t]{0.33\linewidth} 
			\centering
			\includegraphics[width=1.5in, height=1.125in]{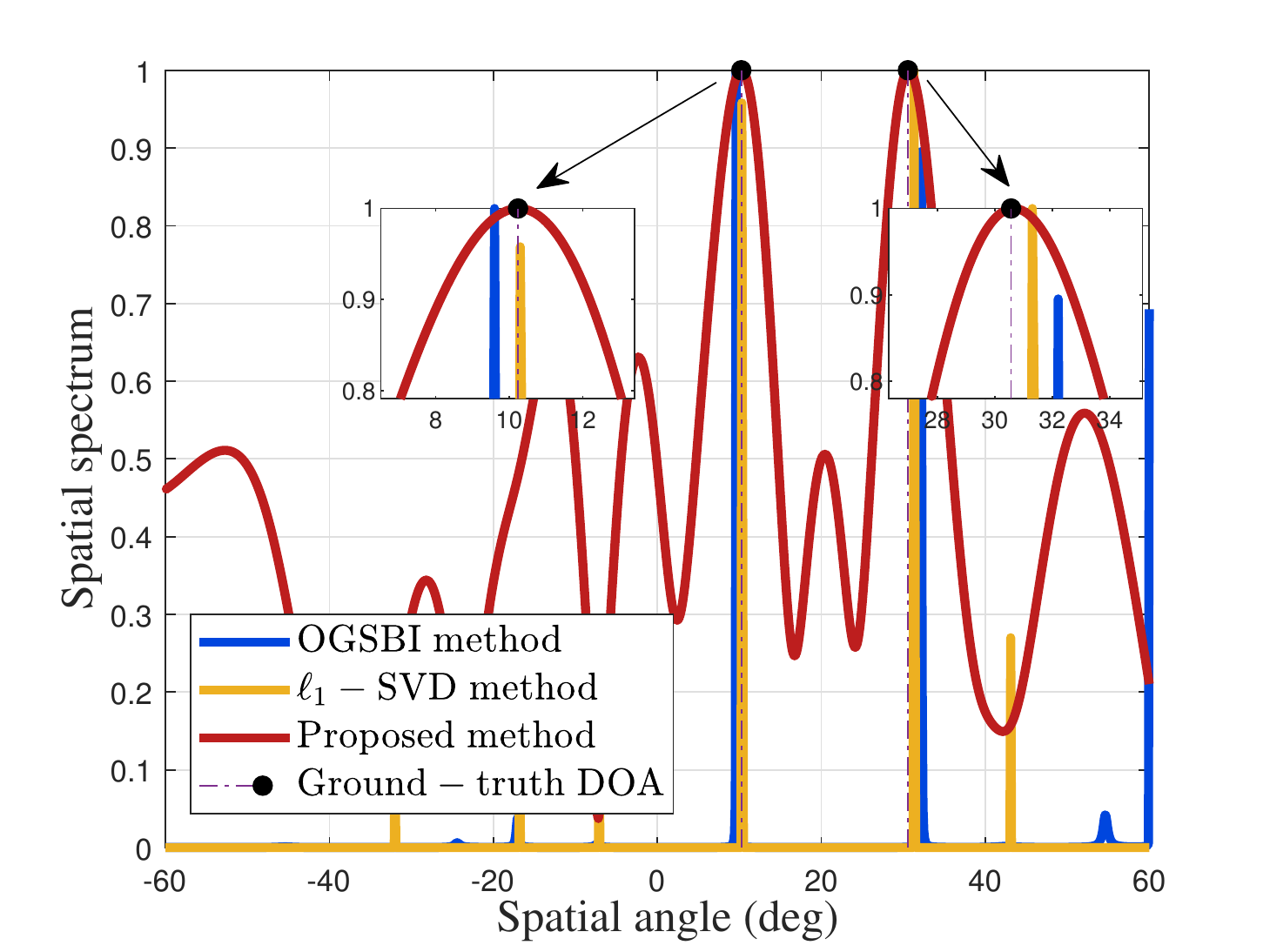} 
			\captionsetup{font={footnotesize}}
			\caption{\,\, Spectrum for DOA estimation.} 
			\label{spectrum} 
		\end{minipage}%
		\begin{minipage}[t]{0.33\linewidth}
			\centering
			\includegraphics[width=1.5in, height=1.125in]{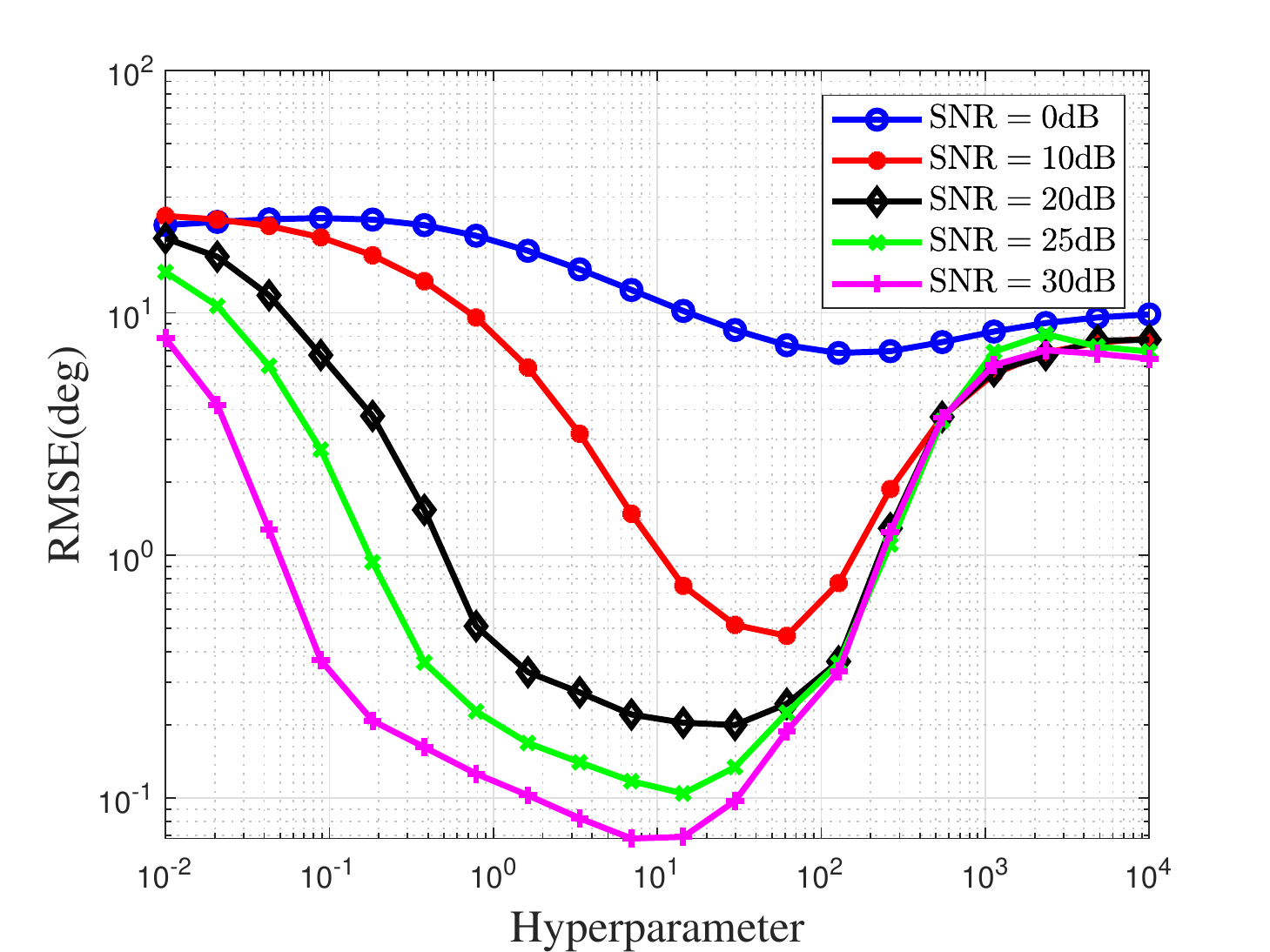}
			\captionsetup{font={footnotesize}}
			\caption{\,\, The RMSE of the DOA esti-\\mation
				with different hyperparameters.}
			\label{hyperparameter}
		\end{minipage}%
		\begin{minipage}[t]{0.33\linewidth}
			\centering
			\includegraphics[width=1.5in, height=1.125in]{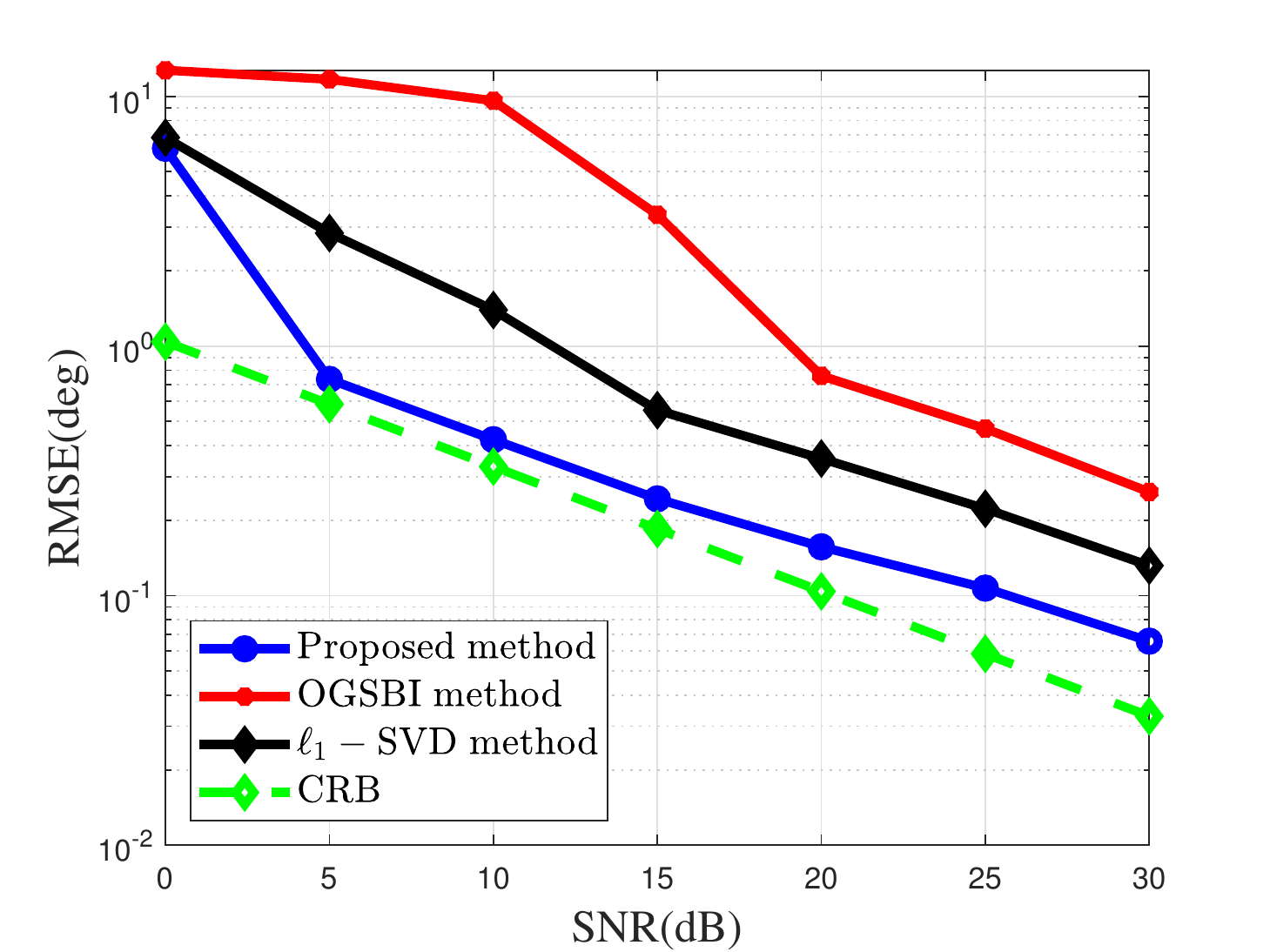}
			\captionsetup{font={footnotesize}}
			\caption{\,\, The RMSE of the DOA estimation \protect\\ with different SNRs.}
			\label{RMSE}
		\end{minipage}
	\end{figure*}	
\begin{table}[H]
	
	\caption{Simulation Settings}
	\label{tab1}
	\centering
	\begin{tabular}{cc}
		\hline
		$\mathbf{Parameter}$             & $\mathbf{Value}$                     \\ \hline
		\rowcolor[HTML]{EFEFEF} 
		N                     & 16                        \\
		P                     & 20                        \\
		\rowcolor[HTML]{EFEFEF} 
		d                     & $\lambda$/2               \\
		K                     & 2                         \\
		\rowcolor[HTML]{EFEFEF} 
		$\boldsymbol{\theta}$ & [$10.24^\circ,30.56^\circ$] \\
		$\theta_{opt}$        & $50^\circ$                 \\ \hline
	\end{tabular}
\end{table}

\begin{figure}[htbp]
		
		\begin{minipage}[t]{0.49\linewidth}
			\centering
			\includegraphics[width=4cm]{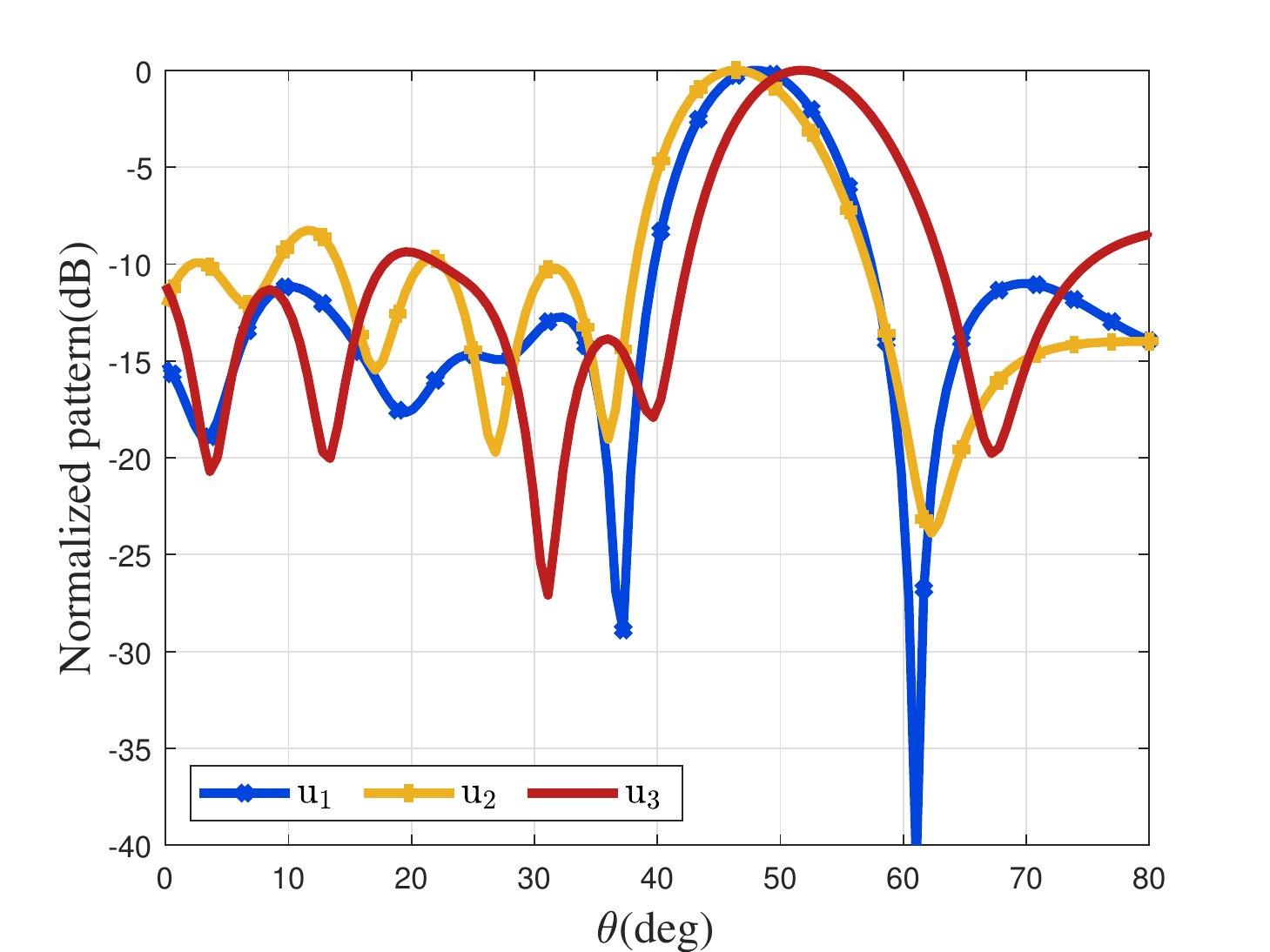}
			\captionsetup{font={footnotesize}}
			\caption{\,\, The beamforming results \\ of the first three measurement vectors.}
			\label{beamforming}
		\end{minipage}
		\begin{minipage}[t]{0.49\linewidth}
			\centering
			\includegraphics[width=4cm]{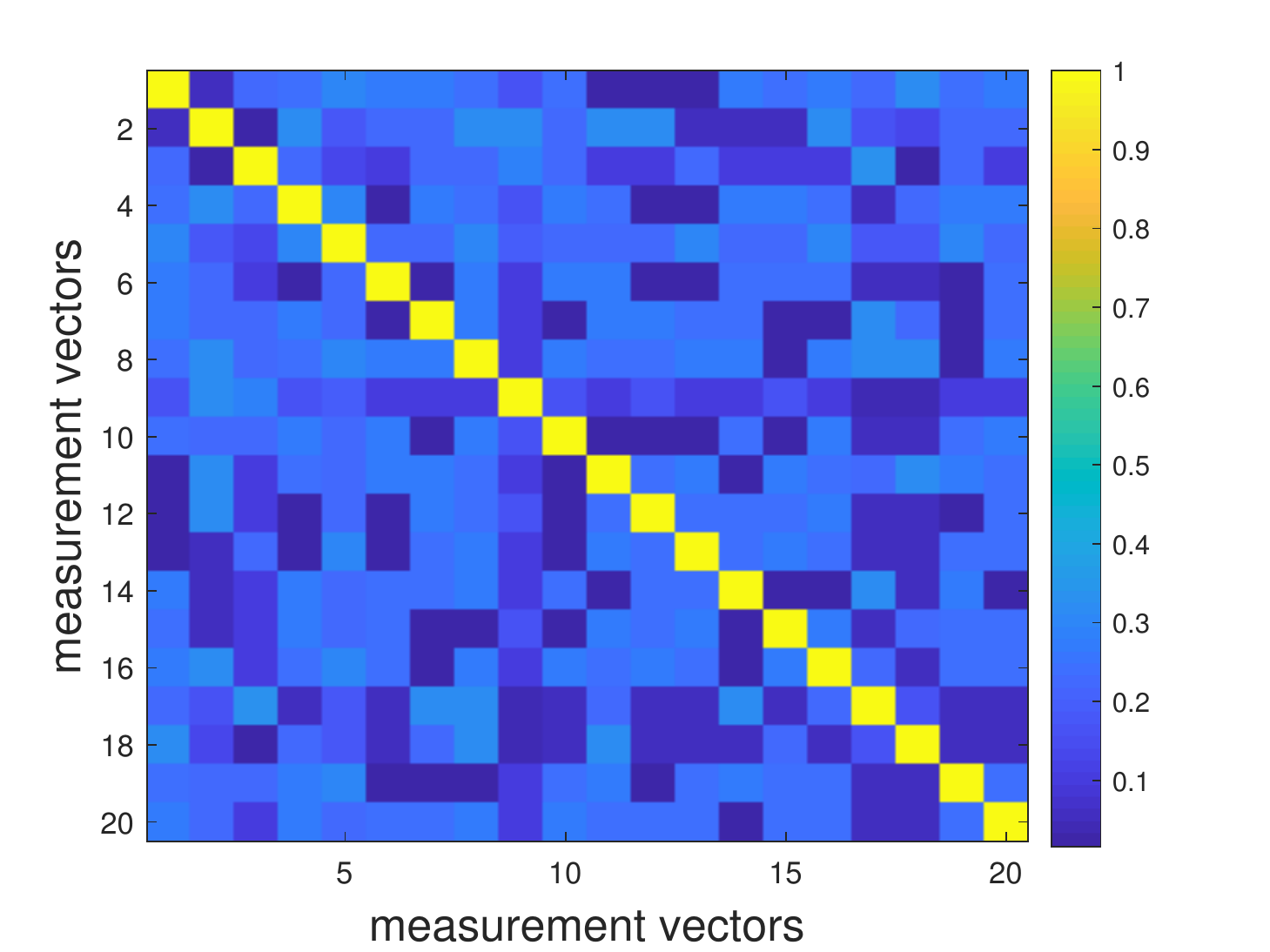}
			\captionsetup{font={footnotesize}}
			\caption{\,\, The correlation coefficients of the measurement matrix.}
			\label{corr}
		\end{minipage}
		
	\end{figure}

	First, the reflection single beam patterns corresponding to the first three measurement vectors are plotted in Fig.~\ref{beamforming}.
	The proposed optimization scheme derives the measurement matrix with mutual coherence lower than 0.35. In Fig.~\ref{corr}, the color of every unit represents the correlation level for its corresponding horizontal and vertical measurement vectors. The darker the color indicates, the lower the correlation. These two figures prove that the proposed optimization scheme in this study can successfully construct the measurement matrix that satisfies both constraints on beam design and mutual coherence.
	
	Then, the spectrum for the DOA estimation is given in Fig.~\ref{spectrum}. The proposed method is compared with two existing methods, including the off-grid sparse Bayesian inference (OGSBI) method \cite{yang2012off}  and the $\ell_1$ singular value decomposition ($\ell_1$-SVD) method \cite{malioutov2005sparse}. These two methods are given a dictionary of the array manifold with $\ang{0.1}$ separation between $\ang{-60}$ to $\ang{60}$. $\ell_1$-SVD method is applied in the proposed system model using $5$ measurements and $4$ snapshots, whose whole measurements number is equal to that of the proposed method as well as the OGSBI method. As shown in Fig.~\ref{spectrum}, the root mean square error (RMSE) of DOA estimation can be obtained. The RMSE is defined as
	\begin{align}
		\text{RMSE}=\sqrt{\frac{1}{MK}\sum_{m=1}^{M}\sum_{k=1}^{K}\left ( \hat{\theta}_{k,m}-\theta_k\right )^2},
	\end{align}
	where M is the number of Monte Carlo trails for RMSE calculation, and $\hat{\theta}_{k,m}$ denotes the estimated DOA of the $k$-th signal source in the $m$-th Monte Carlo trial.
	The RMSEs of the proposed method, the OGSBI method, and the $\ell_1$-SVD method in 20 dB SNR are 0.157, 0.7597, and 0.354 in degree, respectively. The estimation error of the proposed method is much lower than off-grid and on-grid CS methods, which shows the efficiency of the proposed method.

	The regularization parameter $\epsilon$ plays an important role in enhancing the accuracy when estimating DOAs. We conduct an experiment to explore the relation between varied scales of parameters and RMSE. The simulation results are presented in Fig.~\ref{hyperparameter} with the SNR of the received signal being 0~dB, 10~dB, 20~dB, 25~dB, and 30~dB. We can see that estimation accuracy in every SNR level can approach the optimum with a suitable parameter value. In other words, we can derive the expression for the selection of the parameter $\epsilon$ under every SNR as
	\begin{align}
		\epsilon=262.6e^{-0.1327\gamma},
	\end{align}
	where $\gamma$ is SNR in dB. The estimated RMSE of the proposed method compared with other two methods is shown in Fig.~\ref{RMSE} among 100 Monte Carlo trails. It can be inferred that the proposed method outperforms the compared methods.

	\section{Conclusions}\label{sec5}
	We have developed a one-bit RIS-assisted low-cost DOA estimation system in this paper. By introducing the RIS as a phase shift component, we measure the echoes from the UAVs multiple times and steer the beam towards the receiver for more SNR at the receiver. The proposed optimization scheme successfully solves the measurement matrix with two constraints: beamforming parameters and mutual coherence. Then, regarding the measurement matrix $\boldsymbol{U}$ as the prior information for DOA estimation, we utilize the proposed atomic norm-based method to estimate the DOAs. The estimation accuracy of the proposed method reaches optimum by fitting the relation between the parameter and SNR.

	\bibliographystyle{IEEEtran}
	\bibliography{IEEEabrv,reference}
\end{document}